\documentclass[11pt,preprint]{aastex62}

\usepackage{natbib}
\usepackage{float}
\usepackage{graphicx}
\usepackage{amssymb,amsmath,mathtools}

\accepted{ApJ, March 26, 2019}
\shorttitle{LMC and the Hubble Constant}
\shortauthors{Riess et al.}

\newcommand{\xddots}{%
  \raise 5pt \hbox {.}
  \mkern 6mu
  \raise 1pt \hbox {.}
  \mkern 6mu
  \raise -3pt \hbox {.}
}

\newcommand{\bq}{\begin{equation}} 
\newcommand{\eq}{\end{equation}}

\newcommand{\Planckdifflmc}{$  3.6  $}
\newcommand{\Planckdiffmwf}{$  3.9  $}
\newcommand{\Planckdifflmcf}{$  3.7  $}
\newcommand{\Planckdifflmcmw}{$  4.6  $}

\newcommand{\Planckdiff}{$  4.4 \,$}

\newcommand{\holmc}{$  74.22  \pm  1.82  $ km s$^{-1}$ Mpc$^{-1}$ \,}
\newcommand{\holmcnu}{$  74.22  \pm  1.82 $}
\newcommand{\homwfnu}{$  73.94  \pm  1.58 $}
\newcommand{\holmcfnu}{$  73.40  \pm  1.55 $}
\newcommand{\holmcmwnu}{$  74.47  \pm  1.45 $}
\newcommand{\hoallthree}{$  74.03  \pm  1.42  $ km s$^{-1}$ Mpc$^{-1} \,$}
\newcommand{\hoallthreebfnu}{$ {\bf  74.03  \pm  1.42 }$}

\newcommand{\uncallthree}{$  1.91 \% \, $}

\newcommand{\diffv}{0.036 and 0.030 mag \,}
\newcommand{\diffi}{0.018 and 0.036 mag \,}
\newcommand{\diffh}{-0.032 and 0.039 mag \,}
\newcommand{\diffw}{-0.040 and 0.040 mag \,}

\newcommand{\diffvmw}{0.024 mag \,}
\newcommand{\diffimw}{0.038 mag \,}
\newcommand{\diffhmw}{-0.056 mag \,}
\newcommand{\diffwmw}{-0.051 mag \,}

\newcommand{\beq}{\begin{equation}}
\newcommand{\eeq}{\end{equation}}
\newcommand{\beqa}{\begin{eqnarray}}
\newcommand{\eeqa}{\end{eqnarray}}

\newcommand{\PL}{$P\hbox{--}L \,$}

\newcommand{\lcdm}{\hbox{$\Lambda$CDM \,}}

\newcommand{\ndb}{\multicolumn{1}{r}{\textcolor{blue}{$\dots$}}}
\newcommand{\ndr}{\multicolumn{1}{r}{\textcolor{red}{$\dots$}}}
\newcommand{\tcb}{\textcolor{blue}}
\newcommand{\tcr}{\textcolor{red}}

\long\def\check#1{}
\long\def\hide#1{}

\newcommand{\numc}{$ 70 \ $}

\begin{document} 

\title{Large Magellanic Cloud Cepheid Standards Provide a 1\%
  Foundation for the Determination of the Hubble Constant and 
  Stronger Evidence for Physics Beyond $\Lambda$CDM}

\author{Adam G.~Riess}
\affiliation{Space Telescope Science Institute, 3700 San Martin Drive, Baltimore, MD 21218, USA}
\affiliation{Department of Physics and Astronomy, Johns Hopkins University, Baltimore, MD 21218, USA}

\author{Stefano Casertano}
\affiliation{Space Telescope Science Institute, 3700 San Martin Drive, Baltimore, MD 21218, USA}
\affiliation{Department of Physics and Astronomy, Johns Hopkins University, Baltimore, MD 21218, USA}

\author{Wenlong Yuan}
\affiliation{Department of Physics and Astronomy, Johns Hopkins University, Baltimore, MD 21218, USA}

\author{Lucas M.~Macri}
\affiliation{Texas A\&M University, Department of Physics and Astronomy, College Station, TX 77845, USA}

\author{Dan Scolnic}
\affiliation{Duke University, Department of Physics, Durham, NC 27708, USA}

\begin{abstract} 

We present an improved determination of the Hubble constant from {\it Hubble Space Telescope (HST)} observations of \numc long-period Cepheids in the Large Magellanic Cloud.  These were obtained with the same WFC3 photometric system used to measure extragalactic Cepheids in the hosts of Type Ia supernovae. Gyroscopic control of {\it HST} was employed to reduce overheads while collecting a large sample of widely-separated Cepheids.  The Cepheid Period-Luminosity relation provides a zeropoint-independent link with 0.4\% precision between the new 1.2\% geometric distance to the LMC from Detached Eclipsing Binaries (DEBs) measured by \citet{Pietrzynski:2019}  and the luminosity of SNe Ia.  Measurements and analysis of the LMC Cepheids were completed prior to knowledge of the new DEB LMC distance.  Combined with a refined calibration of the count-rate linearity of WFC3-IR with 0.1\% precision \citep{Riess:2019}, these three improved elements together reduce the overall uncertainty in the geometric calibration of the Cepheid distance ladder based on the LMC from 2.5\% to 1.3\%.    Using only the LMC DEBs to calibrate the ladder we find $H_0$=\holmc including systematic uncertainties, 3\% higher than before for this particular anchor.  Combining the LMC DEBs, masers in NGC 4258 and Milky Way parallaxes yields our best estimate: $H_0 = $ \hoallthree, including systematics, an uncertainty of \uncallthree---15\% lower than our best previous result.  Removing any one of these anchors changes $H_0$ by less than 0.7\%. The difference between $H_0$ measured locally and the value inferred from Planck CMB and \lcdm is 6.6$\pm 1.5$ km s$^{-1}$ Mpc$^{-1}$ or \Planckdiff $\sigma$ (P=99.999\% for Gaussian errors) in significance, raising the discrepancy beyond a plausible level of chance.  We summarize independent tests which show this discrepancy is not attributable to an error in any one source or measurement, increasing the odds that it results from a cosmological feature beyond $\Lambda$CDM.
\end{abstract} 


\section{Introduction} 

Cepheid variables in the Magellanic Clouds \citep{Leavitt:1912,Hertzsprung:1913} have long played a starring role in the distance scale and the determination of the present value of the expansion rate of the Universe, the Hubble constant ($H_0$). With knowledge of the distance to the Large Magellanic Cloud (LMC), our nearest Cepheid-rich neighbor, we can directly determine the absolute magnitudes of these pulsating stars.  Cepheids have been  observed with the {\it Hubble Space Telescope (HST)} in the hosts of type Ia supernovae (SNe Ia) at distances of up to 25~Mpc with WFPC2 \citep{freedman01,sandage06} and up to 
40~Mpc with ACS and WFC3 \citep{Riess:2016} to measure the far greater luminosities of these exploding stars. The resulting ability to determine distances to SNe Ia deep into the Hubble-Lema{\^\i}tre flow completes a distance ladder that provides the most precise, model-independent and direct means for determining $H_0$. Knowledge of this cosmological parameter remains central to describing the present state of our Universe and setting expectations for its fate.

Besides characterizing the state of our Universe, refined measurements of $H_0$ may also be pointing towards a new wrinkle in the cosmological model. Measurements of the distance ladder  with improved precision and control of systematics from the SH0ES Team \citep[][hereafter R16]{Riess:2016} demonstrate that the Universe is expanding at present about 9\% faster than inferred from the \lcdm model calibrated by Planck CMB data \citep{Planck:2018} from the early Universe, with a significance of $3.6\sigma$ 
\citep[][hereafter R18a]{Riess:2018b}. The higher local value results from the use of any one of 5 independently determined, geometric distance estimators used to determine the luminosity of Cepheids, including masers in NGC$\,$4258 \citep{humphreys13,Riess:2016}, 8 detached eclipsing binaries (DEBs) in the LMC \citep{Pietrzynski:2013}, and 3 distinct approaches to measuring Milky Way (MW) parallaxes \citep{benedict07}---with the most recent from HST spatial scanning 
\citep[][hereafter R18b]{Riess:2018a} and Gaia DR2 (R18a).  Further out, the distance ratios to SN~Ia hosts provided by Cepheids have been confirmed to a precision of 2-3\% by a dozen measured with the Tip of the Red Giant Branch \citep[TRGB,][]{Jang:2017, Jang:2017a, Hatt:2018a, Hatt:2018b} and Miras \citep{Huang:2018, Yuan:2017}. Strong-lensing results from the H0LiCOW team \citep{Birrer:2018} are fully independent of all rungs of the distance ladder yet find a similar value of $H_0$ from the late Universe, one which is 2.3 $\sigma$ ($P=98\%$) higher than the Planck-calibrated value.  At the other end of time, the low value of $H_0$ predicted from the early Universe is corroborated by independent measurements of the CMB or $\Omega_B$ with BAO data \citep{Addison:2017} and from ``inverse distance ladders'' such as the one built by the Dark Energy Survey and which is calibrated from the sound horizon at z$\sim$1000 \citep[][see Discussion for further consideration of these results]{Macaulay:2018}.  With multiple, independent checks now established at both ends of cosmic history, this ``$H_0$ Tension'' between the early and late Universe, as it is widely known, may be interpreted as evidence for a new cosmological feature such as exotic dark energy, a new relativistic particle, dark matter-radiation or neutrino-neutrino interactions, dark matter decay, or a small curvature, each producing a different-sized shift \citep{Renk:2017,Khosravi:2017,Aylor:2018,Di-Valentino:2018,Mortsell:2018,DEramo:2018,Kreisch:2019,Barenboim:2019,Pandey:2019,Vattis:2019} with some proposals spanning the full discrepancy while improving the agreement between the model and CMB data \citep{Poulin:2018}. Pinpointing the cause of the tension requires further improvement in the local measurements, with continued focus on precision, accuracy, and experimental design to control systematics.

New measurements of 20 late-type DEBs the LMC from \citet{Pietrzynski:2019} offer the most precise foundation to date to geometrically calibrate this distance ladder.  The current approach to measuring these geometric distances uses long-baseline near-infrared interferometry of individual late-type giants to measure their angular sizes and derives a purely empirical relation between their surface brightness and color (SBCR) with a scatter of only 0.018 mag.  That such a relation exists is a direct consequence of the Planck Law.  Applying this relation to a late-type giant in a DEB system yields the geometrically-calibrated angular diameter of the star from its color and brightness.  Combined with the physical radius derived from radial velocities and eclipsing light curves yields a purely geometric distance with a typical precision of $\sim$ 2\% per system.  These DEB measurements appear quite robust, as the variance of the sample is fully characterized by the method and there is no dependence on astrophysical models; the details can be found in \citet{Pietrzynski:2013} and \citet{Pietrzynski:2019}. The most recent result uses an improved SBCR whose calibration otherwise systematically limits the measurement and an expanded sample of 20 DEBs to measure the distance to the center of the LMC to 1.2\% (0.0263 mag) precision.
	
To fully exploit the improved precision of the LMC distance, we need greater control of systematic errors than past measurements.  Simply comparing the brightnesses of Cepheids in the LMC to those in SN~Ia hosts measured from two different telescopes would incur a net systematic 2-3\% error, just from the use of different photometric systems with their individual zeropoint uncertainties.  Measurements of Cepheids in the NIR are especially important in order to mitigate systematic errors from extinction and metallicity variations.  Yet photometric errors are larger in the NIR, as ground-based bandpasses are unstable and are redefined nightly by the atmosphere.  Even for the best calibrated ground-based system in the NIR, 2MASS, the {\it systematic} uncertainty in the transformation to the best-match WFC3 {\it F160W} band (after accounting for bandpass differences) is found empirically to be $\sigma \approx 0.03-0.04$~mag \citep{riess11b}. This is not surprising, as the {\it absolute} zeropoints of each are (claimed to be) known to only 0.02--0.03 mag \citep{Skrutskie:2006, Kalirai:2009}. Thus, mixing ground-based photometry from Cepheids in the NIR from \cite[][hereafter M15]{Macri:2015}  or \citet{persson04} with those from HST in SN~Ia hosts incurs a $\sim$ 1.4-1.8\% systematic error in distance measurements, swamping the improved LMC distance precision {\it unless a single, stable system is used to measure both sets of Cepheids and nullify zeropoint errors}.  Even using a single photometric system, it is necessary to calibrate its ability to measure relative fluxes across the 10 mag range between Cepheids in the LMC and SN~Ia hosts. Fortunately, concurrent work has now calibrated the linearity of the WFC3-IR detector to a precision of 2.4~mmag across this range, making the higher precision sought feasible \citep{Narayan:2018,  Calamida:2018, Riess:2019, Bohlin:2019}.

In the past, the prospect of observing many Cepheids in the LMC directly with HST was hampered by the high cost of observatory  overheads. Because the LMC is nearby, its Cepheids are far apart in angle, and thus observing each with {\it HST} required a dedicated pointing (with attendant guide star acquisition overhead).  However, using a newly available commanding and control sequence under purely gyroscopic-control called ``DASH'' \citep[Drift And SHift;][]{Momcheva:2017} we observed up to a dozen LMC Cepheids in 3 filters in a single orbit, obtaining HST photometry for 70 widely-separated LMC Cepheids with WFC3 in 3 filters.  This photometry establishes a new, zeropoint-independent link between LMC Cepheids and those in the hosts of SNe Ia.
	 
In \S 2 we present the observations and measurements of these LMC Cepheid standards, their {\PL} relations in \S 3 and the impact on the Hubble constant in \S 4.  We consider systematics related to the determination of $H_0$ in \S 5.  We point out that all the photometric  measurements of the LMC Cepheids presented here were completed and this manuscript finalized in advance of the availability of the new LMC distance, its uncertainty and the revised geometry of the LMC based on the new DEBs measurements in \citet{Pietrzynski:2019}.  After this became available, only the final determination of $H_0$ was completed.

\section{LMC Cepheid Standards}

\subsection {DASHing through the LMC} 

The \numc LMC Cepheids presented here were imaged in three bands, two in the optical with WFC3-UVIS ($F555W$ and $F814W$) and one in the near-infrared with WFC3-IR ($F160W$), in two HST programs: GO-14648 and GO-15146 (PI: Riess).  All data frames are available in MAST\footnote{MAST is the archive site for STScI and HST at {\tt /https://mast.stsci.edu}}.  The observations were taken between 2017-01-09 and 2018-12-16 and are identified and described in Table 1.

Measuring the mean magnitudes of a large number of Cepheids in the LMC with the narrow-angle instruments on HST poses unique challenges. The mean separation of $P>6$ day LMC Cepheids is $\sim$10 \arcmin\ ($\sim$15 \arcmin\ for $P>10$ day Cepheids), well in excess of the full 2\arcmin\ WFC3 field of view,  so in almost all cases only a single Cepheid can be observed per image. Although the necessary exposures times are only a few seconds, with normal observing procedures, each Cepheid observation requires a new FGS guide star acquisition with an overhead of 6 minutes, by far the longest interval in the observation. In addition, WFC3 can only hold one full UVIS frame in memory before requiring a memory buffer dump which requires 350 seconds.  Thus, full-frame imaging of LMC Cepheids with short exposures is extremely inefficient and time consuming; given the demand for the use of {\it HST},  such observations are unlikely to be undertaken.

However, we can observe these Cepheids far more efficiently by using the new DASH mode of observing, available since Cycle 24 (2016), which uses the HST gyroscopes for both slewing and guiding.  This mode is highly efficient for our short integrations of 2-2.5 seconds; during this time, the smearing of the PSF from the expected gyro drift of 1--2 mas/s is $\sim$ 0.003\arcsec\, which is negligible compared to the 0.04\arcsec\ size of pixels in WFC3-UVIS or the 0.128\arcsec\ pixels for WFC3-IR, thereby saving the overhead of repeated guide star acquisitions.  By selecting groupings of Cepheids, typically within less than a degree, it is possible to slew {\it HST} 5--10\arcmin\ between successive Cepheids with an overhead of 2 minutes per slew, observe each on a subarray in the near-IR, flip the channel select mechanism no more than once per orbit (an observatory requirement), reverse the path, and observe each target again in two optical filters (also in subarray mode), collecting up to 12 Cepheids in 3 filters in one orbit--and without exceeding the memory buffer.  This strategy requires that the accumulated pointing error during the orbit remain smaller than half the subarray size ($\sim 10\arcsec$), which is consistent with the aforementioned {\it expected} gyro drift under normal conditions. Ground-based light curves can be used to adjust each single-epoch magnitude to the epoch of mean intensity as done in R18a and R18b.  By definition, these phase corrections are zeropoint-independent since they are calculated relative to the average magnitudes of each Cepheid (R18).

Our first attempt at implementing this observing sequence, visit 8 of GO-14648, succeeded as planned on 2017 Jan 1, collecting the images of 12 LMC Cepheids spread over 0.5 degrees in 3 bands, 36 exposures in all, obtained within a single HST orbit (just over one hour of elapsed time).  The accumulated drift from the commanded position did not exceed 6\arcsec, and we observed a mean drift of $1.2$ mas/s (see Figure 1); thus all Cepheids landed well inside their chosen apertures.  We selected a sample of 100 LMC Cepheids from the sample of M15, assigning greater priority to Cepheids with longer periods (better analogues of those in SN~Ia hosts) and those requiring shorter slews.

Unfortunately, the start of this program and its follow-on, GO-15146, coincided with a period of {\it severe} degradation in the performance of HST Gyro 2, which caused its typical drift rate to exceed its nominal value by more than an order of magnitude.  The specific  magnitude and direction of the gyro drift at any point in time  was found to be highly unpredictable. To overcome increasing gyro drift rates with time, we redesigned our subsequent observations to use larger subarrays and thus provide greater margin for erratic gyro slewing.  The larger frames required more on board memory for storage and in this state we could observe $\sim$ 6-10 Cepheids before filling the memory.  By mid-2018 Gyro 2 drifts increased so that the accumulated pointing errors under gyro control reached $20-70\arcsec$ after 2000 seconds (see Figure 1), in some cases exceeding the radius of the full WFC3 frame and thus ensuring the target would miss the field regardless of the array size used.  A few targets were subsequently observed under FGS control to complete the intended 3 bands of imaging.  We concluded the observing program with 70 completed targets each with 3 colors, somewhat less than the 100 targets expected with nominal gyro performance, but representing only a net 17\% loss in the statistical power of the sample.   Gyro 2 stopped operating at the end of this program, and was replaced in the operational chain by Gyro 3, which may have elevated noise; it is unclear how this mode will perform in the new gyro configuration.  Should HST drop to fewer than three gyro control, the feasibility of DASH-mode observations is unlikely.

The erratic gyro slewing also made it challenging to determine exactly where HST was pointed in each subarray and thus where the target star was located (under gyro control, HST cannot use the astrometry of a known guide star to establish the pointing coordinates).  We developed and employed an algorithm for matching the apparent positions of stars to a catalogue of the LMC to identify the location of each target.  The X and Y pixel positions of each Cepheid in the observation as they actually occurred {are} given in Table 1.
  
\subsection {Photometry}

We measured the photometry of the Cepheids using small apertures with radii of 3 pixels for WFC3-UVIS and WFC3-IR to reduce source contamination (from cosmic rays or nearby stars) and to minimize sky noise.  Aperture photometry has other advantages over PSF fitting for this application, including lower systematics if the inner PSF varies (which could potentially result from gyro drifts) and less variation from PSF undersampling in single frames.  In practice we did not find any measurable variation in the size or shape of the PSF due to gyro drifts. During the period of degraded Gyro 2 performance, the measured mean drift rate for targets landing in the arrays was $5$~mas/s (with a peak of $13$~mas/s).  The corresponding mean drift over the 2.5 second exposure was 0.3 pixels for WFC3-UVIS and 0.1 pixels for WFC3-IR, well within the aperture and compensated by the use of an aperture correction between $r=3$ and $r=10$ pixels derived from the mean of all Cepheid PSFs.

Measurements were made on the fully calibrated frames from the MAST archive, using the CTE-corrected frames for WFC3-UVIS images.  Each calibrated image was multiplied by the pixel area map, a necessary step to  obtain correct point-source photometry for an image flat-fielded to  constant flux per unit area.  For WFC3-UVIS data, we derived and applied an aperture correction from 3 to 10 pixel radius ($0.12\arcsec$ to $0.4\arcsec$) in order to utilize the 10-pixel zeropoint  provided by STScI for each UVIS CCD.  Our subarrays all used CCD2, for which the adopted zeropoints (magnitude of a star which produces  1 $e^- s^{-1}$) are 25.727 mag for $F555W$ and 24.581 mag for $F814W$ (Vega system).  This procedure matches the calibration used for optical Cepheid photometry in R16 and \citet{Hoffmann:2016} and the application to the distance scale remains independent of the accuracy of these zeropoints as long as the same consistent value is used to measure all Cepheids along the distance ladder.

For WFC3-IR $F160W$ we used aperture corrections of 0.063~mag from a radius of 10 pixels to infinity (provided by STScI) or equivalently 0.200~mag from a 3 pixel radius aperture (measured by us).  We adopted a zeropoint of 24.71 mag at infinite radius (Vega system), derived to yield the same mean photometry whether measured from  these apertures on the original frames or from PSF fitting on  resampled images, which is the methodology employed by R16 and \citet[][hereafter R11]{riess11} for SN Ia host images using a scale of 0.08\arcsec\  pixel$^{-1}$ and a flux drop fraction of 0.6. R16 and R11  used images of the {\sc Calspec} reference star P330 as the  reference for the shape and scale of the PSF; this is also  one of the stars used to set the zeropoint by STScI and has a similar color to Cepheids.  As a  result, the zeropoint we derived to ensure uniformity of  Cepheid photometry matches the official STScI result to 0.01 mag. By comparing photometry measured with apertures on the original  pixel scale and PSF fitting on the resampled images used by R16, we estimate a systematic uncertainty between measurement techniques of 3~mmag.  

We also found a small systematic difference in the WFC3-UVIS photometry  of Shutter A vs. Shutter B images for these very short exposures. \citet{Sahu:2014} established that Shutter B causes extra instrument vibrations and can affect the PSF and move some flux outside the aperture  for very short exposure times.  We determined and corrected for a  difference of $\pm 6$~mmag in $F555W$ and $\pm 3.5$~mmag in $F814W$ depending on which shutter was used.
 
Next, we applied a correction for the expected difference between the magnitude of each each Cepheid at the observed phase and the magnitude at the epoch of mean intensity of its light curve.  These phase corrections are derived from ground-based light curves of each Cepheid in filters with wavelengths best corresponding to the WFC3 filters. Because the phase corrections are relative quantities, they do not change the zeropoint of the light curves, which remain on the {\it HST} WFC3 natural system \footnote{In practice the ground-based light curves are transformed to the {\it HST} system using color terms.  While an uncertainty in color term could produce systematic errors, these are negligible.  We determined empirically that a 10\% error in the color terms changes the mean phase correction by $\leq$ 0.1 mmag.}.  We derived and applied these phase corrections following the same methodology described in R18b. The periods and phases for $F555W$ and $F814W$ were determined using the {\it V}- and {\it I}-band light curves from OGLE surveys \citep{2005AcA....55...43S,2008AcA....58...69U,2015AcA....65....1U}. For a few Cepheids (OGL0434, OGL0501, OGL0510, OGL0512, OGL0528, OGL0545, OGL0590, OGL0712, OGL0757, OGL0966, OGL0992) we also included {\it V}-band light curves from the ASAS survey \citep{1997AcA....47..467P} and/or ASAS-SN survey \citep{2014ApJ...788...48S,2017PASP..129j4502K} to increase the baseline coverage. We made use of the {\it J}- and {\it H}-band light curves from M15 and \citet{persson04} to correct the $F160W$ random phased measurements to mean intensity. The standard deviations of these corrections are 0.29, 0.17 and 0.11~mag in $F555W$, $F814W$, and $F160W$, respectively, decreasing with the smaller light curve amplitudes at redder wavelengths.  Phase corrections also account for the difference between the Cepheid light curve magnitude mean (the average of many measured magnitudes) and the magnitude at the epoch of mean intensity (the standard convention for distance measurements).  This expected difference is consistent with our sample average correction of -0.048, -0.013 and -0.001~mag, in $F555W$, $F814W$ and $F160W$, respectively.   The uncertainties in these phase corrections depends on the quality of the ground-based light curves; the average uncertainty is 0.013, 0.008, and 0.029~mag per epoch in $F555W$, $F814W$, and $F160W$, respectively, which dominates over the statistical photometry errors (i.e., photon statistics) in a single epoch.  The differences between repeat measurements for the same target, available for a subset of 19 epochs and filters, is consistent with these uncertainties. The final mean individual uncertainty for these 70 Cepheids is 0.016, 0.012, and 0.029~mag in $F555W$, $F814W$ and $F160W$, respectively.  The final mean photometry for each Cepheid in 3 colors is given in Table 2.

For distance measurements and for the determination of H$_0$, it is useful to combine these three bands into the same single, reddening-free Wesenheit index \citep{madore82} used by R16 for measuring extragalactic Cepheids in SN~Ia hosts: \bq m^W_H=m_{F160W}-0.386(m_{F555W}-m_{F814W}). \eq  where the value of 0.386 is derived from the reddening law of \citep{Fitzpatrick:1999} with $R_V=3.3$.  R16  and \cite{Follin:2017} considered a broader range of reddening laws.

\noindent where the coefficient of the color term is derived from the reddening law, accounting for the relatively low absorption in the near-infrared.  Using the Wesenheit magnitude thus defined has the additional benefit of tightening the resultant {\PL} relation, since across the instability strip, intrinsically redder Cepheids are fainter.  For a {\PL} that is well-sampled across the intrinsic color range, the difference in the color ratio for dust and intrinsic color has no impact on relative distance measurements, since the intrinsic component cancels out when comparing the \PL relations of different hosts.  The $m^W_H$ values for our 70 targets have an individual mean uncertainty of 0.030 mag, including photometric measurement errors, phase corrections, and error propagation to the Wesenheit magnitude.

We can also compare the \PL relations we obtain from HST photometry with ground-based results, by transforming the ground ($V$,$I$,$H$) magnitudes from M15 to the HST natural system ($F555W$, $F814W$, $F160W$) via the equations given in R16.  We emphasize that this transformation is primarily for comparison purposes, and is not required in the calibration of the Cepheid \PL in the HST photometric system.  The results are shown in Figure~2, where we excluded three Cepheids, OGLE0712, OGLE1539 and OGLE1677, whose HST images reveal a relatively bright and nearby star ($\Delta$mag$<3$, $\Delta r<1\arcsec$) that would significantly contaminate ground-based photometry of the Cepheid for typical ground-based image quality. We find mean differences (in the direction Ground$-$ {\it HST\/}) and sample dispersions (SDs) of \diffv in $F555W$, \diffi in $F814W$, \diffh in $F160W$ and \diffw in $m^W_H$.  There are a couple of outliers in the comparison between space and the ground (two out of 67 in each filter); these are marked in Figure~2. These mean differences in zeropoints are similar to those found when comparing ground and {\it HST} magnitudes of Milky Way Cepheids in \citet{Riess:2018b}: \diffvmw for $F555W$, \diffimw for $F814W$, \diffhmw for $F160W$, and \diffwmw for $m^W_H$.  While the dispersion between these ground and HST zeropoints for this LMC sample is 0.04 mag, the dispersion {\it between} these differences for the LMC and Milky Way is $<0.02$~mag, suggesting these HST-to-ground zeropoint {\it differences} are primarily systematic. The presence of systematic errors in zeropoints from either space or ground facilities reinforces the value of maintaining a single, stable photometric system to nullify these errors along the distance ladder.
   
A comparison of near-infrared photometry of Cepheids along the distance ladder, even within the same photometric system, must also account for another systematic difference which appears when measuring bright and faint sources with HgCdTe detectors.  This effect, called Count-Rate Non-Linearity (CRNL) or reciprocity failure, is different from the more commonly  considered non-linearity of  {\it total} measured counts (often approaching saturation) which is already corrected in the MAST pipeline through calibrations determined from varying the length of integrations.  In contrast, CRNL causes  photons at low count rates to be detected or collected less efficiently than photons at high count rate, regardless of the length of the exposure.
     
Recently, \cite{Riess:2019} derived a more precise characterization of the CRNL of WFC3 through a combination of comparisons of cluster star photometry between WFC3-IR and WFC3-UVIS at overlapping wavelengths and by comparing observed and synthetic magnitudes of white dwarfs \citep{Calamida:2018, Narayan:2018} and by further extending the results to brighter count rates.  Combining these results with previous measurements and those from the WFC3 grism \citep{Bohlin:2019} provides a consistent and improved characterization of the CRNL of WFC3-IR, of $0.75\pm0.06$\% per dex, with no apparent wavelength dependence, now measured across 16 astronomical magnitudes.  This improves by a factor of 4 on the  previous measurement of $1.00\pm0.25$\% per dex \citep{Riess:2010}.
     
The correction for CRNL is more properly (i.e., physically) considered as accounting for dimming of fainter sources (e.g., Cepheids in SN~Ia hosts) relative to brighter sources (such as HST reference stars) as a result of charge trapping.  However, because calculations involving the distance ladder are only sensitive to {\it the difference} in measured flux and because the photometry of the faint Cepheids in SN Ia hosts in R16 did not (by convention) include a correction for CRNL, we account for the net difference here.  We have therefore added to the $m^W_H$ photometry in Table 2 the $0.0300\pm 0.0024$ mag mean correction to the the bright LMC Cepheids to account for the 4.0 dex flux ratio in $F160W$ between these LMC Cepheids and the sky-dominated Cepheids observed in SN~Ia hosts and NGC 4258.  This is the same convention used for the MW Cepheids measured with {\it HST} and presented in R18.
  
Lastly, due to the inclination of the LMC, we expect some Cepheids to be closer or farther than average by a few hundreths of magnitude in distance modulus, depending on their projected distance from the LMC line of nodes.  \citet{Pietrzynski:2019} used the DEBs to constrain a tilted plane geometry for the LMC with inclination $i=25.0^\circ \pm 4$ and position angle of the line of nodes $\Theta=132^\circ \pm 10$ for a center of mass from \citet{van-der-Marel:2014} of $\alpha_0=80.05^\circ$, $\delta_0=-69.30^\circ$.  The mean correction for the DEBs to the center was 0.001 mag.  For consistency with the geometry of the DEBs from \citet{Pietrzynski:2019} we use the same model to correct the Cepheids to the center and mean DEB.  Because more of our Cepheids are South-West of the line of nodes (which is tilted away from us) our mean Cepheid, according to the model, is more distant than the center by 0.011 mag (SD$= 0.010$ mag), with a full range of 0.004 mag closer to 0.038 mag more distant.  The values of $m^W_H$ in Table 2 include these individual corrections to account for the projected distance of the Cepheids from the line of nodes. To evaluate the systematic uncertainty in this correction we consider other geometries.  \citet{van-der-Marel:2014} used kinematic constraints to derive an inclination of $i=26.2^\circ$ and position angle of the line of nodes $\Theta=154.5^\circ$ which produced a mean correction for our Cepheids of 0.013 mag.   An alternative model derived from the positions and fluxes of 2000 LMC Cepheids by \citet{Nikolaev:2004} with $\alpha_0=79.4^\circ$, $\delta_0=-69.03^\circ$, $i=30.7^\circ$ and $\Theta=151.0^\circ$ yields a mean correction of 0.016 mag.   We will consider the standard deviation of these 3 models, 0.002 mag, to be a systematic uncertainty associated with the geometry of the LMC.  As we will show, this term is subdominant to other error terms.

\startlongtable
\begin{deluxetable*}{cccccccc}
\tabletypesize{\scriptsize}
\tablewidth{0pt}
\tablenum{1}
\tablecaption{Observations of LMC Cepheids (Complete table available in published, electronic version)\label{tb:phot}}
\tablehead{\colhead{Cepheid} &  \colhead{frame} & \colhead{filter} & \colhead{MJD$^*$} & \colhead{exptime (s)} & \colhead{array} & \colhead{X} & \colhead{Y}}
\startdata
OGL0434 & 
idi532e4q & F555W &  1357.849  & 2.00  &  UVIS &      608 &       515 \\
OGL0434 & 
idi532e5q & F814W &  1357.850  & 2.00  &  UVIS &      613 &       509 \\
OGL0434 & 
idi518iaq & F160W & 1176.417  & 2.22  & IRSUB256 &      138 &      136  \\
OGL0434 & 
idi532dqq & F160W & 1357.776  & 2.22  & IRSUB256 &      138 &      139  \\
\hline
\enddata
\tablecomments{$^*$ MJD-57000.0 }
\end{deluxetable*}

\startlongtable
\begin{deluxetable*}{ccccccccccccc}
\tabletypesize{\scriptsize}
\tablewidth{0pt}
\tablenum{2}
\tablecaption{Photometric Results LMC Cepheids (Complete table available in published, electronic version)\label{tb:phot2}}
\tablehead{\colhead{Cepheid} &  \colhead{RA} & \colhead{DEC} & \colhead{Geo} & \colhead{log Period} & \colhead{$F555W$} & \colhead{$\sigma$}   & \colhead{$F814W$} & \colhead{$\sigma$}  & \colhead{$F160W^a$} & \colhead{$\sigma$}  & \colhead{$m^{W,b}_H$} & \colhead{$\sigma$}}
\startdata
OGL0434 & 74.114583  & -69.379611  & 0.028  & 1.482  & 13.131  & 0.028  & 12.208  & 0.011  & 11.321  & 0.018  & 10.966  & 0.021   \\
OGL0501 & 74.462625  & -69.958250  & 0.034  & 1.367  & 13.623  & 0.022  & 12.693  & 0.012  & 11.770  & 0.021  & 11.406  & 0.023   \\
OGL0510 & 74.523208  & -69.454333  & 0.027  & 1.566  & 13.457  & 0.037  & 12.299  & 0.021  & 11.232  & 0.042  & 10.787  & 0.045   \\
OGL0512 & 74.545000  & -69.949694  & 0.033  & 1.595  & 13.134  & 0.025  & 12.005  & 0.017  & 11.038  & 0.017  & 10.598  & 0.020   \\
OGL0528 & 74.636583  & -70.346028  & 0.038  & 1.553  & 13.175  & 0.052  & 12.156  & 0.021  & 11.226  & 0.020  & 10.824  & 0.029   \\
OGL0545 & 74.696292  & -70.061583  & 0.034  & 1.199  & 14.414  & 0.045  & 13.349  & 0.010  & 12.311  & 0.018  & 11.895  & 0.025   \\
OGL0590 & 74.921417  & -69.456111  & 0.025  & 1.502  & 13.470  & 0.025  & 12.382  & 0.014  & 11.311  & 0.038  & 10.895  & 0.039   \\
\hline
\enddata
\tablecomments{$^a$Does not include addition of $0.0300 \pm 0.0024$ mag to correct CRNL 4 dex between MW and extragalactic Cepheids.}
\tablecomments{$^b$Includes subtraction of geometric correction to LMC line of nodes and addition of $0.0300 \pm 0.0024$ mag to correct CRNL 4 dex between MW and extragalactic Cepheids.}
\end{deluxetable*}

\section {Period-Luminosity Relations}

In Figure 3 we show the relations between Cepheid period and magnitudes in $F555W$, $F814W$, and $F160W$ as well as two Wesenheit indices,  $ m^W_H=m_{F160W}-0.386(m_{F555W}-m_{F814W}) $ and  $ m^W_I=m_{F814W}-1.14(m_{F555W}-m_{F814W}) $.   The slopes of these relations (Table 3) match well those derived from larger ground-based samples of LMC Cepheids from \citet{Soszynski:2008} and M15.  We determined the intercepts and scatter for $m^W_H$ from a $2.7\sigma$ clipped mean (this threshold is derived from Chauvenet's criterion where we expect 0.5 outliers at $>$ 2.7 $\sigma$ from a normal distribution with $N=70$ Cepheids).  
The $m^W_H$ relation gives the lowest scatter, SD$=0.075$ mag, with only two marginal outliers, OGL0992 and OGL0847, at 2.8 and $3.1\sigma$, respectively (including both of these outliers results in a slightly increased $SD=0.086$).  For the same set of Cepheids, their ground-based magnitudes transformed to $m^W_H$ give a somewhat higher scatter of 0.084 (or 0.103 with no outliers removed); the increased scatter may be due in part to occasional light contamination from nearby stars in the lower quality images from the ground. To determine the {\it intrinsic} scatter in the HST-based relations we subtract the estimated measurement errors, finding a result of 0.069 mag (equivalent to 3.2\% in distance) for a single Cepheid; which is at or near defining the lowest apparent dispersion of any known sample of Cepheids.

The scatter, after removing the same two outliers as above, is seen to decline with increasing wavelength, as may be expected due to the reduced impact of differential extinction and the narrower intrinsic width of the {\PL} relation.  Two of the Cepheids are particularly red, with $F555W-F814W$ of 1.3 mag and 1.6 mag for OGL1945 and OGL1940, respectively, or more than 3$\sigma$ redder than the mean ($<F555W-F814W>=1.00$, SD$=0.09$ without these two), but neither are outliers in the reddening-free $m^W_H$.   

Although the Cepheid ground sample from M15 is an order of magnitude larger in size than the one observed with {\it HST}, the latter is more heavily weighted towards longer period Cepheids which offers advantages when comparing to Cepheids in SN Ia hosts.  The {\it HST} LMC sample has a mean period of 16 days ($\log P=1.21$) and 45 Cepheids with $P>10$ days while the M15 ground sample has a mean period of 6.7 days ($\log P=0.83$) and 109 Cepheids with $P>10$ days.   The mean Cepheid period in the SN Ia hosts is $\log P \sim$1.5 (R16) so the difference in sample mean $\log P$ for the {\it HST} sample is less than half the ground sample.  Thus an uncertainty in the slope of the Cepheid \PL relation will propagate less than half the error in $H_0$ from the {\it HST} sample as from the ground sample.

The formal error in the LMC Cepheid sample mean $m^W_H$ is 0.0092 mag, equivalent to 0.42\% in distance.  The total formal uncertainty for the Cepheid LMC calibration is obtained by combining this uncertainty with the 0.0263 mag (or 1.20\% in distance) error from the DEBs \citep{Pietrzynski:2019}, the 0.0024 mag systematic uncertainty in the CRNL correction, the 0.003 mag difference between methods of measuring photometry, and the 0.002 mag uncertainty due to the LMC geometry.  Together these yield a total uncertainty of 1.28\% for the geometric calibration of the Cepheid distance ladder and $H_0$, which is the smallest error for any Cepheid calibration to date (see Table 4).  Additional uncertainty when comparing these Cepheids to those with different metallicity or mean period will be considered in the next section.

\begin{table}[h]
\begin{small}
\begin{center}
\vspace{0.4cm}
\begin{tabular}{cclc}
\multicolumn{4}{c}{{\bf Table 3:} Period Mean-magnitude Relations from HST LMC Cepheids}\\

\hline
\hline

band &  slope$^*$ & intercept$^c$ & scatter$^c$ \\
\hline
$F555W$ & -2.76 & 17.638 & 0.312  \\
$F814W$ & -2.96 & 16.854 & 0.202  \\
$F160W$ & -3.20 & 16.209 & 0.104  \\
$m^{W,b}_I$ & -3.31 & 15.935 & 0.085  \\
$m^W_H$ & -3.26 & 15.898$^a$ & 0.075  \\
\hline
\hline
\multicolumn{4}{l}{$^*$ from \citet{Soszynski:2008}, M15 and R16}\\
\multicolumn{4}{l}{$^a$ includes CRNL}\\
\multicolumn{4}{l}{$^b$ $R=A_I/(A_V-A_I)=1.3$ from \citet{cardelli89} $R_V=3.1$ as \cite{freedman01}}\\
\multicolumn{4}{l}{$^c$ outliers are OGL0992 and OGL0847}\\
\end{tabular}
\end{center}
\end{small}
\end{table}

Previous studies \citep[][and references therein]{ Bhardwaj:2016} have suggested a change in slope of the {\PL} at $P=10$ days at optical wavelengths, though its significance has been marginal and debated.  As in R16, our primary fit for $H_0$ in the next section allows for a change in slope at $P=10$ days; if the change in slope is real, not allowing for this degree of freedom in the solution could otherwise introduce a systematic error when comparing samples of Cepheids with differing mean periods. However, R16 found no evidence of a break in the $m^W_H$ relation with slopes of $-3.25\pm 0.02$ and $-3.26\pm 0.02$~mag/dex below and above $P=10$ days, respectively, averaged across all extragalactic Cepheids (including the 785 LMC Cepheids from M15).  For only the 70 LMC Cepheids studied here there is a slope difference of $\Delta=0.20\pm0.30$~mag/dex across $P=10$ days with the measured slope at $P>10$ days ($N=43$) of $-3.38\pm0.07$~mag/dex, which is $2.1\sigma$ steeper than the R16 mean of extragalactic Cepheids.  Because the sample is relatively small, the slope is sensitive to the low number of rare long period Cepheids. \citet{persson04} measured a similar number ($N=39$) of $P>10$ day LMC Cepheids from the ground in the $H$-band and found a slope of $-3.16\pm 0.10$~mag/dex;  they included two Cepheids at $P=99$ and 134 days, neither of which was observed here and which pull the slope to lower values. The opposite has been seen for Cepheids in M31, with the longer period end being $\sim 2\sigma$ shallower than the shorter period end, though the combination produces no evidence of a break to 0.02~mag uncertainty \citep[R16 and][]{kodric15}.  As in R16, allowing for a break slightly reduces $H_0$ by 0.4\% and is included, with other variants pertaining to the fitting of the \PL relation, in the systematic error.

\begin{table}[h]
\begin{small}
\begin{center}
\vspace{0.4cm}
\begin{tabular}{lll}
\multicolumn{3}{l}{{\bf Table 4:} Systematic Error Budget for the LMC Cepheid Distance Ladder Calibration} \\
\hline
\hline
error & value & source \\
\hline
LMC {\PL}  Mean & 0.42\% & measured here \\
DEB Mean & 1.20\% & \citet{Pietrzynski:2019} \\
CRNL across 4 dex & 0.11 \% &  \citet{Riess:2019} \\
LMC Geometry & 0.09\% & std dev of 3 geometries (see text) \\
Photometry methods & 0.14\% & measured here\\
\hline
Total & 1.28\% &  \\
\hline
\hline
\end{tabular}
\end{center}
\end{small}
\end{table}

\section {The Hubble Constant}

Following the distance ladder and additional constraints provided in R16, we can use this LMC Cepheid {\PL} relation to help calibrate the luminosity of SNe~Ia and determine the value of $H_0$.  Among the 3 geometric sources previously used by R16 for this purpose, masers in NGC 4258, Milky Way parallaxes, and LMC DEBs, the latter yielded the lowest individual value of $H_0$, $72.04\pm 2.67$~km s$^{-1}$ Mpc$^{-1}$.  This LMC-derived calibration employed 785 Cepheids observed from the ground from M15 and the 8 DEBs from \citet{Pietrzynski:2013}.  R16 assumed a systematic uncertainty between the ground and HST-based zeropoint of $\sigma$=0.03 mag, an estimate in good agreement with the empirical result of a 0.04 mag ground-to-space difference found in \S 2.  This relative zeropoint error limited the available precision well above the apparent error in the mean distance of the LMC Cepheid ground sample of 0.08 mag / $\sqrt{785}$ or 0.003 mag, an order of magnitude lower than the zeropoint uncertainty.  This uncertainty was also significant compared to the previous LMC distance uncertainty from \citet{Pietrzynski:2013} of 0.045 mag, with 0.054 mag (2.5\%) of the combined overall error pertaining to the prior LMC calibration route.

Here we use both the ground and HST sample of LMC Cepheids together, as each provides an important and complementary constraint.  The 70 LMC Cepheids observed with HST alone constrain the error in the mean relative to Cepheids observed by HST in SN~Ia hosts to a precision of just under 0.01 mag.   After including the 0.0263 mag uncertainty of the DEB-based distance from \citet{Pietrzynski:2019} and the uncertainty from the CRNL the total error becomes 0.028 mag (1.28\% in distance; see Table 4)  about half the uncertainty of the LMC combination used in R16.  The much larger ground-measured sample is still very valuable, considered simultaneously, to constrain the {\it slope} of the \PL.  The slope was constrained in R16 from the ground-based sample to $\sigma \sim 0.01$~mag/dex (or $\sigma \sim 0.02$~mag/dex for two slopes if a break was allowed), which is independent of its $\sigma=0.03$~mag zeropoint uncertainty.  Fitting the distance ladder with the system of equations given in R16 and retaining the systematic zeropoint uncertainty for only the ground-based sample optimally leverages both samples. For each Cepheid with a ground and space-based measurement, we include a covariance term equal to the square of the intrinsic LMC dispersion measured here of 0.07 mag to account for this correlated error.  

For the ground-based sample, we include the differences in projected distance to the line of nodes using the model of \citet{Pietrzynski:2019} and their mean LMC distance of $\mu=18.477 \pm 0.0263$~mag based on the DEBs which have already been corrected to the line of nodes.  For all LMC Cepheids we assume a mean [Fe/H]$=-0.30$~dex, which is chosen to be between the mean of -0.33 dex from 22 objects observed spectroscopically by \citet{Romaniello:2008} and -0.27 dex which is the mean of the photometric metallicity map of \citet{Choudhury:2016} at the positions of the Cepheids from M15.  This is slightly different than the value of -0.25 dex adopted by R16.  Following \citet{Anderson:2018}, we have also included a small correction of 0.0074 mag for the additional mean flux statistically and {\it physically associated with Cepheids} that is not resolved at the distances of the SN~Ia hosts but which is resolved in the closer LMC as discussed in \S 4.  This is addition to the use in R16 of artificial star measurements to account for mean additional light due to {\it chance superposition} on crowded backgrounds.

\begin{deluxetable}{lrr}
\tablewidth{0pc}
\tablecaption{Best Estimates of H$_0$ Including Systematics\label{tb:h0}}
\tablenum{5}
\tabletypesize{\small}
\tablewidth{0pc}
\tablehead{\multicolumn{1}{l}{Anchor(s)} & \colhead{Value} & \multicolumn{1}{r}{$\Delta$ Planck$^*$+}\\
\multicolumn{2}{r}{[km s$^{-1}$ Mpc$^{-1}$]} & \lcdm ($\sigma$)}

\startdata
LMC  & \holmcnu  & \Planckdifflmc \\
\hline
\multicolumn{2}{l}{Two anchors} \\
\hline
LMC + NGC$\,$4258 & \holmcfnu & \Planckdifflmcf \\
LMC + MW & \holmcmwnu & \Planckdifflmcmw \\
NGC$\,$4258 + MW & \homwfnu & \Planckdiffmwf \\
\hline
\multicolumn{2}{l}{\bf Three anchors (preferred)} \\
\hline
{\bf NGC$\,$4258 + MW + LMC} & \hoallthreebfnu & \Planckdiff \\
\hline
\enddata
\tablecomments{$^*: H_0=67.4\pm0.5$ km s$^{-1}$ Mpc$^{-1}$\\ \citep{Planck:2018}}
\end{deluxetable}

Using only the LMC distance from \citet{Pietrzynski:2019} to geometrically calibrate the Cepheid luminosities, we find \holmc including its systematic uncertainty calculated by the analysis variants method given in R16.  The value is higher than the value of $72.04 \pm 2.67$ km s$^{-1}$ Mpc$^{-1}$ from R16 by 2.2 km s$^{-1}$ Mpc$^{-1}$ (about 0.8 $\sigma$ or 2.9\%) due primarily to the 1\% decrease in the LMC distance between \citet{Pietrzynski:2013} and \citet{Pietrzynski:2019} (which increases $H_0$ by 1\%) and a $\sim 2$\% increase from the use of the HST photometric system for the Cepheids.  The overall uncertainty in $H_0$ using the geometric LMC calibrations has declined by 40\%, 25\% due to the improved LMC distance and 15\% due to the use of a single photometric system which nullifies the relative zeropoint uncertainty.   Because the LMC Cepheids have a lower metallicity than those in SN Ia hosts (or the MW or NGC 4258), there is an additional uncertainty of 0.9\% when using the LMC as an anchor due to the uncertainty in the empirically constrained luminosity-metallicity relation.  These improvements together make the LMC, with a 2.4\% total uncertainty in $H_0$ comparable in precision (actually better) as an anchor of the distance ladder to the use of all MW Cepheid parallaxes, with the masers in NGC 4258 somewhat lower at 3.4\% uncertainty; all are individually consistent within $1.3\sigma$ in terms of their independent geometric distance uncertainties.

In Table 5 we also list the result of combining the LMC with the MW Cepheid parallaxes (R18a), with the masers in NGC 4258, and every combination of using only a pair of anchors.  These two-anchor (or one-anchor-out) combinations now have a smaller range of 1.07 km s$^{-1}$ Mpc$^{-1}$ (compared to 2.42 km s$^{-1}$ Mpc$^{-1}$ in R16) due to the increase in the result from the LMC.  Indeed, leaving out any one of three anchors by choice, which is a reasonable test of robustness, changes $H_0$ by only $\sim$0.5 km sec$^{-1}$ Mpc$^{-1}$ or $<$ 0.7\%.  For those inclined to disfavor any one anchor, these combinations offer a best result without the influence of the given anchor.  

However, the best and preferred result comes from including all three anchors, giving \hoallthree, a total uncertainty of \uncallthree including systematics. Compared to the {\it predicted} value of $H_0$ of $67.4 \pm 0.5 $ km s$^{-1}$ Mpc$^{-1}$ from the \citet{Planck:2018} CMB data in concert with the cosmological model, \lcdm, this measurement differs by \Planckdiff $\sigma$ (P=99.999\%). 

In Table 6 and Figure 5 we give a detailed breakdown of all sources of uncertainty in the determination of $H_0$ here and compared to R16.  The primary changes between the present uncertainties in $H_0$ and those in R16 result from improvements in the anchor measurements from the LMC and Milky Way. The contributed uncertainty from Milky Way Cepheid parallaxes has decreased from 2.5\% to 1.7\% due to new parallax measurements from HST spatial scanning (R18b) and from Gaia Data Release 2 (R18a) and from the use of WFC3 to measure their photometry on the same photometric system as Cepheids in SN Ia hosts.  These improvements in the Milky Way anchor alone reduced the overall uncertainty in $H_0$ from 2.4\% to 2.2\% (R18a).  An even greater improvement in the LMC anchor is now realized, decreasing its contributed uncertainty from 2.6\% to 1.5\%.  While there is a small increase in uncertainty in the \PL intercept due to the smaller sample of LMC Cepheids here, this is more than offset by the smaller systematic uncertainty in their photometric zeropoint.  We also note that there is an {\it increase} in the overall uncertainty due to the relation between Cepheid metallicity and luminosity.  The metallicity term we derived from our analysis of all Cepheid data (R16) is -0.17 $\pm 0.06$ mag per dex, similar to \citet{Gieren:2018} who finds -0.22 mag per dex in the NIR for a lower range of metallicity.  The product of the mean, sub-solar metallicity for the LMC Cepheids and the uncertainty in this term is 0.9 \%.  The other two anchors have Cepheids with near solar metallicities that are much closer to those in the SN hosts so the overall uncertainty in $H_0$ due to metallicity is weighted down by these anchors to 0.5\%.  

\begin{deluxetable}{llrrrrrrr}
\tablecaption{Recent H$_0$ Error Budgets (\%)}
\tabletypesize{\small}
\tablewidth{0pc}
\tablenum{6}
\tablehead{\colhead{Term} & \colhead{Description} & \multicolumn{3}{c}{\tcb{Riess+ (2016)}} & \multicolumn{3}{c}{\tcr{Here}} \\
\colhead{} & \colhead{} & \multicolumn{1}{c}{\tcb{\scriptsize LMC}} & \multicolumn{1}{c}{\tcb{\scriptsize{MW}}} & \multicolumn{1}{c}{\tcb{\scriptsize 4258}} & \multicolumn{1}{c}{\tcr{\scriptsize LMC}} & \multicolumn{1}{c}{\tcr{\scriptsize{MW}}} & \multicolumn{1}{c}{\tcr{\scriptsize 4258}}}
\startdata
$\sigma_{\mu, \rm anchor}$  &   Anchor distance  & \tcb{2.1} & \tcb{2.1} & \tcb{2.6} & \tcr{1.2} & \tcr{1.5} & \tcr{2.6}\\
$\sigma_{{\rm PL, anchor}}$  &  Mean of {\PL} in anchor & \tcb{0.1} & \ndb &\tcb{1.5} & \tcr{0.4} & \ndr & \tcr{1.5} \\
$R \sigma_{\lambda,1,2}$  & zeropoints, anchor-to-hosts & \tcb{1.4} & \tcb{1.4} & \tcb{0.0} & \tcr{0.1} & \tcr{0.7} & \tcr{0.0} \\
$\sigma_{Z}$  & Cepheid metallicity, anchor-hosts & \tcb{0.8} & \tcb{0.2} & \tcb{0.2} & \tcr{0.9} & \tcr{0.2} & \tcr{0.2} \\
\tableline
$ $  & subtotal per anchor & \tcb{2.6} & \tcb{2.5} & \tcb{3.0} & \tcr{1.5} & \tcr{1.7} & \tcr{3.0} \\
                     &                             & \multicolumn{3}{p{2.6cm}}{\raisebox{.66\baselineskip}{$\underbrace{\hspace{2.6cm}}$}} & \multicolumn{3}{p{2.6cm}}{\raisebox{.66\baselineskip}{$\underbrace{\hspace{2.6cm}}$}} \\
\multicolumn{2}{l}{All Anchor subtotal}  & & \tcb{1.6} & & & \tcr{1.0} & \\
\tableline
$\sigma_{{\rm PL}}/\sqrt{n}$  &  Mean of {\PL} in SN~Ia hosts & & \tcb{0.4} & & & \tcr{0.4} & \\
$\sigma_{\rm SN}/\sqrt{n}$  &  Mean of SN~Ia calibrators (\# SN) & \multicolumn{3}{l}{\ \ \ \ \ \ \ \ \ \tcb{1.3 (19)}}&\multicolumn{3}{l}{\ \ \ \ \ \ $\,\,$\ \ \tcr{\bf 1.3 (19)}}\\
$\sigma_{m-z}$  &  SN~Ia $m$--$z$ relation & & \tcb{0.4} & & & \tcr{0.4} & \\
$\sigma_{\rm PL}$ & {\PL} slope, $\Delta$log\,$P$, anchor-hosts & & \tcb{0.6} & & & \tcr{0.3} & \\
\tableline
\multicolumn{2}{l}{statistical error, $\sigma_{{\rm H}_0}$}  & & \tcb{2.2} & & & \tcr{1.8} & \\
\tableline
\multicolumn{2}{l}{Analysis systematics$^a$} & & \tcb{0.8} & & & \tcr{0.6} & \\
\tableline
\multicolumn{2}{l}{{\bf Total uncertainty on} $\sigma_{{\rm H}_0}$ [\%]} & & \tcb{2.4} & & &\tcr{\bf 1.9} & \\
\tableline
\enddata
\tablecomments{$^a$ Systematic errors calculated as standard deviation of 23 analysis variants presented in R16, given here as 1.48*median[abs(variants-mean(variants))].}
\end{deluxetable}

\section{Discussion}

\subsection{Systematics: Cepheid Associated Flux} 
The photometric measurements of Cepheids from R16 in SN~Ia hosts and NGC 4258 account for the mean additional light due to chance superposition on crowded backgrounds through the use of artificial star measurements.  However, the possibility of light from stars which are {\it physically} associated with the Cepheids and unresolved at their distances for SN Ia hosts (5-40 Mpc) but which is resolved in the LMC at 50 kpc (or the Milky Way at 2-3 kpc) and thus excluded from measurement would have a differential effect that could bias the determination of $H_0$. \cite{Anderson:2018} quantified this ``associated-light bias'' by studying its two plausible sources, wide binaries ($a_{rel}>400$ AU) and open clusters (closer binaries are unresolved in all cases).  They found that the mean effect of wide binaries was negligible (0.004\% in $H_0$) because Cepheids dominate companions in luminosity.  Closer binaries, while more common, are unresolved in either anchor galaxies or SN hosts so even the tiny contamination of Cepheid flux from a companion, $\sim$0.02\% in distance, cancels along the distance ladder due to its presence for all Cepheids (assuming binarity is common in all hosts).  To quantify the impact of open clusters, they analyzed the regions around a large sample of Cepheids in M31, 450 Cepheids with UV HST imaging from the PHAT program \citep{Dalcanton:2012}.  They found 2.4\% of Cepheids are in such clusters and that the photometric bias averaging over Cepheids in or out of clusters is 0.0074 mag for $m^W_H$.  This value might be considered an upper limit to the bias since there is also a ``discovery bias'' to exclude even the small fraction of Cepheids in bright clusters from a distant sample.  The additional constant flux which is unresolved for distant Cepheids in clusters would decrease the amplitude of Cepheid light curves.  \citet{Anderson:2018} found a mean bias for a Cepheid in a cluster in M31 of 0.30 mag in $m^W_H$ corresponds to a bias of 0.8 mag at visual wavelengths, near or brighter than the limit of 0.5 mag contamination that \citet{Ferrarese:2000} determined would preclude discovery of a Cepheid due to the flattening of its light curve.  In the other direction one might posit a somewhat larger clustered fraction in SN~Ia hosts than in M31 (M31 being somehow unusual) but this direction is limited by the greater ages of Cepheids ($30-300$~Myr) than clusters with only $\sim10$\% of massive embedded clusters surviving for more than 10 Myr \citep[][and sources within]{Anderson:2018}.  Indeed, M31 provides the best analog for the SN~Ia hosts (high metallicity spiral) for which an up-close, external view of Cepheid environments is available.  Such accounting for the MW may await improved parallaxes.  In this regard the LMC is unusual, with a greater frequency of Cepheids in clusters and a higher concentration of massive clusters (likely due to its high rate of recent star formation), with 7.2\% of $P>10$ day Cepheids in clusters (with fewer than 4 Cepheids per cluster).  The LMC harbors two Cepheid-rich clusters each with 24 Cepheids, eight times the number of Cepheids as the richest MW cluster.  Due to the great resolution of {\it HST} in the LMC, this excess of clusters around Cepheids in the LMC has no photometric impact on the measurement of $H_0$.  Here we have included the expected impact of such flux based on the example of M31 and conclude it does not produce an impediment to measuring $H_0$ to 1\%. Higher resolution imaging in the future from JWST or large ground-based telescopes with adaptive optics may allow for measuring the fraction of Cepheids in clusters in more distant galaxies.

\subsection{Prospects for reducing the $H_0$ uncertainty: SN Statistics and Systematics}

As shown in Table 6 and Figure 5, with the new LMC anchor measurements in hand, the single largest source of uncertainty in $H_0$ now lies in determining the mean luminosity of the SNIa calibrators.   Future reductions in the uncertainty in $H_0$ will require increasing the number of these calibrations.  The sample in R16 had 19 calibrators, which were chosen based on SNIa light-curve quality requirements and for their presence in nearby ($z \leq 0.01$), late-type, globally star forming, non-edge on hosts which would thus be expected to yield a good sample of Cepheids.  New observations with {\it HST} by the SH0ES program are underway which will double this sample to 38 calibrators, making the sample complete to $z \sim 0.011$, and will reduce the uncertainty in this term by $\sqrt{2}$ and the likely overall error on $H_0$ to $\sim$ 1.5\%.

With the prospect of reaching sub-percent uncertainty in the mean of SNIa calibrators, even greater attention must be given to controlling potential systematic uncertainties in the measurement of SN Ia distances.  One path to both better standardize SN~Ia brightnesses and control systematics has been to search for additional observables that may correlate with distance residuals beyond light-curve shapes \citep{Phillips:1993} and colors \citep{Riess:1996, Phillips:1999, Tripp:1998}.  As the statistical leverage of larger and better calibrated SN samples have grown \citep{Betoule:2014,Scolnic:2018}, evidence has arisen of an environmental parameter that appears to correlate with SN distance residuals, albeit at a lower level than the preceding SN parameters.   The most widely used form for this environmental parameter is to use the host galaxy mass as done by nearly all recent cosmology analyses including those from the SNLS \citep{sullivan11}, the Joint SNLS-SDSS Light Curve Analysis \citep[JLA;][]{Betoule:2014}, Pan-Starrs \citep{Rest:2014,Scolnic:2014,Scolnic:2018,Jones:2018a},  SCP \citep{Suzuki:2012}, Dark  Energy Survey \citep{DES-Collaboration:2018},  SH0ES (R11, R16), and CSP \citep{Burns:2018}.  More recent studies have found a weaker \citep{Scolnic:2018} or absent \citep{DES-Collaboration:2018} environmental dependance compared to past studies \citep{hicken09b,kelly10,lampeitl10,Sullivan:2010}, likely due to improved accounting for SN selection bias \citep{Kessler:2017}. Likewise, recent measurements of $H_0$ by R16 and \citet{Burns:2018} have followed this approach and corrected for the empirical dependence of host mass in their samples.

We can evaluate the systematic error resulting from SN Hubble residual dependence on a different environmental parameter such as star formation rate, colors, metallicity and population ages.  An environmental parameter can impact the determination of the Hubble constant if two conditions are met: 1) the parameter has a {\it significant} relationship with the distance residuals for the {\it same} SN sample and method of distance determination used to measure the Hubble constant and 2) there is a mean difference in the parameter between the SNe in Cepheid hosts and those used to calibrate the Hubble expansion.  \citet{Jones:2018} looked for additional environmental dependencies in the same SN sample and distance fitting method used by R16 to determine $H_0$ (the only such study of the R16 sample) including local and global color (in the UV to address star formation), local and global mass, and local specific star formation (suggested by \citet{Rigault:2018} to relate to environmental age).  They found none of these are significant ($<$ 2 $\sigma$) in the R16 distance residuals (which already include a correction for host mass and SN selection bias) with the possible exception of local mass ($\sim 3 \sigma$).  Significance aside, \citet{Jones:2018} combined the sample differences and size of the dependencies  and for different environmental parameters they found an impact on $H_0$ of 0.3\% to 0.5\%.   \citet{Roman:2018} also found no significant relation  between Hubble residuals and local SN UV color for 123 SNe Ia at $z<0.1$ (1.7$\sigma$ without a host mass correction).  In a study of SDSS SNe Ia, \citep{Rose:2019} performed a principal component analysis to create the optimal combination of multiple additional parameters that could correlate with distance residuals finding an additional correlation between the inferred environmental age and Hubble residuals with 2.1$\sigma$ significance which would produce a $\sim$0.4\% change to $H_0$ (after accounting for the present host mass correction which is included in R16) if the same correlation exists in the nearby SNe used in R16.  These {\it residual} environmental dependencies are much smaller than the present overall uncertainty, of low significance, and if true, would remain subdominant to the future statistical uncertainty in $H_0$ with a larger sample size of SN~Ia calibrators.

The safest path to avoid possible (even unknown) environmental systematics would be to eliminate {\it the potential cause} for a mean difference in environments between the SNe in Cepheid hosts and those used to calibrate the Hubble expansion by employing the same criteria to select SNe in the two relevant samples.   A simple approach would be to limit SNe in the Hubble Flow sample to those in {\it globally} late-type, star forming hosts. These are the same criteria used to select likely Cepheid hosts; and because the nature of the {\it local} SN environment is not relevant and thus not a factor in this selection, the statistics of local environments would be similar in both samples.  R16 employed both a late-type only and globally star forming only selection as a variant for the determination of $H_0$ and demonstrated a change of $H_0 < 0.3\%$. 

The advantage of this approach is illustrated in the study of H-$\alpha$ local to the SN site.  In studies of the unpublished SN Factory sample \citet{Rigault:2018,Rigault:2013} suggested the presence of local H-$\alpha$, a consequence of a very young population, may have an important dependence on SN distance with strong H-$\alpha$ (for a specific local mass) producing fainter SNe (a similar effect as having a low mass host).  Because the Hubble flow sample may include SNe with early-type hosts with lower local H-$\alpha$, a higher mean local H-$\alpha$ may be expected in the purely late-type (and globally star forming) hosts of the Cepheid calibrator sample.  Thus, {\it if} a local H-$\alpha$ relation existed in the SN sample used to determine $H_0$, it could raise its value.  However, including only late-type hosted SNe in the Hubble flow sample would remove the underlying cause of such a bias.  \citet{Anderson:2015a} provides measurements of the relative strength of H-$\alpha$ at the sites of 98 SNe Ia in exclusively late-type, star forming hosts, including 20 of the 38 selected for Cepheid measurements, which can be used to test this approach.  Of the selected Cepheid hosts, 14 of 20 (70\%) (or 8 of 11 from R16) have no detected H-$\alpha$ (at the 0\% level relative to the host) while in the non-Cepheid selected sample, 43 of 78 (55\%) show no H-$\alpha$.  The mean of the Cepheid sample is thus consistent with albeit {\it lower} in H-$\alpha$, demonstrating that limiting the samples to the same host type removes the source of a sample mean difference of H-$\alpha$ beyond that following statistical fluctuations (or another factor unrelated to selection) and the potential for biasing the value of $H_0$ whether a dependence exists in the R16 Hubble flow sample (which has {\it not} been shown) or not.  In this case, an even balance of SNe with high, local H-$\alpha$ among late-type, star forming hosts is expected since the presence of SN local H-$\alpha$ never enters their selection. Limiting the Hubble flow sample to the same criteria as Cepheid hosts reduces the sample by less than half and thus has little impact on the precision of $H_0$.   

\subsection{Present Status}

The recent landscape of model-independent, direct measurements of $H_0$ from the late Universe and predictions from the early Universe is shown in Figure 4.  While it is difficult and perhaps debatable to identify the precise threshold at which a tension passes the point of being attributable to a fluke, the one presently involving $H_0$ appears to have passed that point \footnote{In the search for a new particle where an unexpected ``bump'' in events may occur at any point along a continuum of energies, the ``Look Elsewhere'' effect of having too many chances (bins) for a random fluctuation to appear is typically overcome with a high threshold for significance, usually 5$\sigma$.  Here where we have only one cosmological parameter (or at most a few) to compare between the early and late Universe, and only one Universe to observe, fewer throws of the dice are possible, so we consider the present discrepancy quite significant.}.

The higher, local value of H$_0$ from the distance ladder has been determined through five independent geometric absolute calibrations of the Cepheid \PL relation including new Milky Way parallaxes from {\it HST} spatial scans (R18b) and from Gaia DR2 (R18a).   As shown in Table 5, any one of the 3 types of anchors can be removed with a change in $H_0$ of $<$ 0.7\%, much smaller than the present 9\% discrepancy.  The relative distances from Cepheids match those from the Tip of the Red Giant Branch \citep[TRGB,][]{Jang:2017,Jang:2017a,Hatt:2018b,Hatt:2018a} and Miras \citep{Huang:2018, Yuan:2017} to a mean precision of 2-3\%. Strong-lensing results from the HOLiCOW team have corroborated the local value of H$_0$ independent of the distance ladder \citep{Birrer:2018}.  At the other end of cosmic time, the low expected value of H$_0$ predicted from the early Universe has been confirmed by different sources of CMB and BAO data and by an independent calibration of BAO from measurements of $\Omega_B$ and knowledge of the CMB temperature \citep{Addison:2017}.  It is important to note that ``inverse distance ladders'' and calibrations of BAO or SNe originating from the sound horizon ($z\sim1000$) or from CMB measurements all fall into this ``early Universe'' category of $H_0$ inference. With multiple, independent corroborations now demonstrated at both ends of cosmic history, we may need to seek resolution in a refinement of the model that joins them, (Vanilla) $\Lambda$CDM.

A new feature in the dark sector of the Universe appears increasingly necessary to explain the present difference in views of expansion from the beginning to the present.  The general considerations by \citet{Aylor:2018} argue for an injection of energy or expansion preceding recombination which would shrink the sound horizon inferred from the CMB and used to calibrate BAO with a specific scenario offered by \citet{Poulin:2018}.  The dark radiation sector (i.e., neutrinos) also provides a possible source for alleviation of the $H_0$ difference through interactions of additional components \citep{Kreisch:2019}.

Continued pursuit in precision in the determination of $H_0$ is also needed to transition from the discovery of a difference to a diagnosis of its source.  Additional observations of giants and pulsating stars in more hosts of SNe~Ia are underway and should further refine $H_0$. Less predictable but highly sought are contributions from gravitational wave sources as standard sirens \citep{Schutz:1986,Abbott:2017,Chen:2018}.  Improvements in parallaxes from future Gaia data releases are also expected to continue to increase the precision of the distance ladder in the near term.

\bigskip

\bigskip

\acknowledgements

We are grateful to Roeland van der Marel for assistance with the geometry of the LMC.

Support for this work was provided by the National Aeronautics and Space Administration (NASA) through programs GO-14648, 15146 from the Space Telescope Science Institute (STScI), which is operated by AURA, Inc., under NASA contract NAS 5-26555. A.G.R., S.C. and L.M.M. gratefully acknowledge support by the Munich Institute for Astro- and Particle Physics (MIAPP) of the DFG cluster of excellence ``Origin and Structure of the Universe.''

This research is based primarily on observations with the NASA/ESA {\it Hubble Space Telescope}, obtained at STScI, which is operated by AURA, Inc., under NASA contract NAS 5-26555.

The {\it HST} data used in this paper are available as part of the MAST archive which can be accessed at {\tt http://archive.stsci.edu/hst/}.

\begin{figure}[ht]
\vspace*{150mm}
\figurenum{1}
\includegraphics{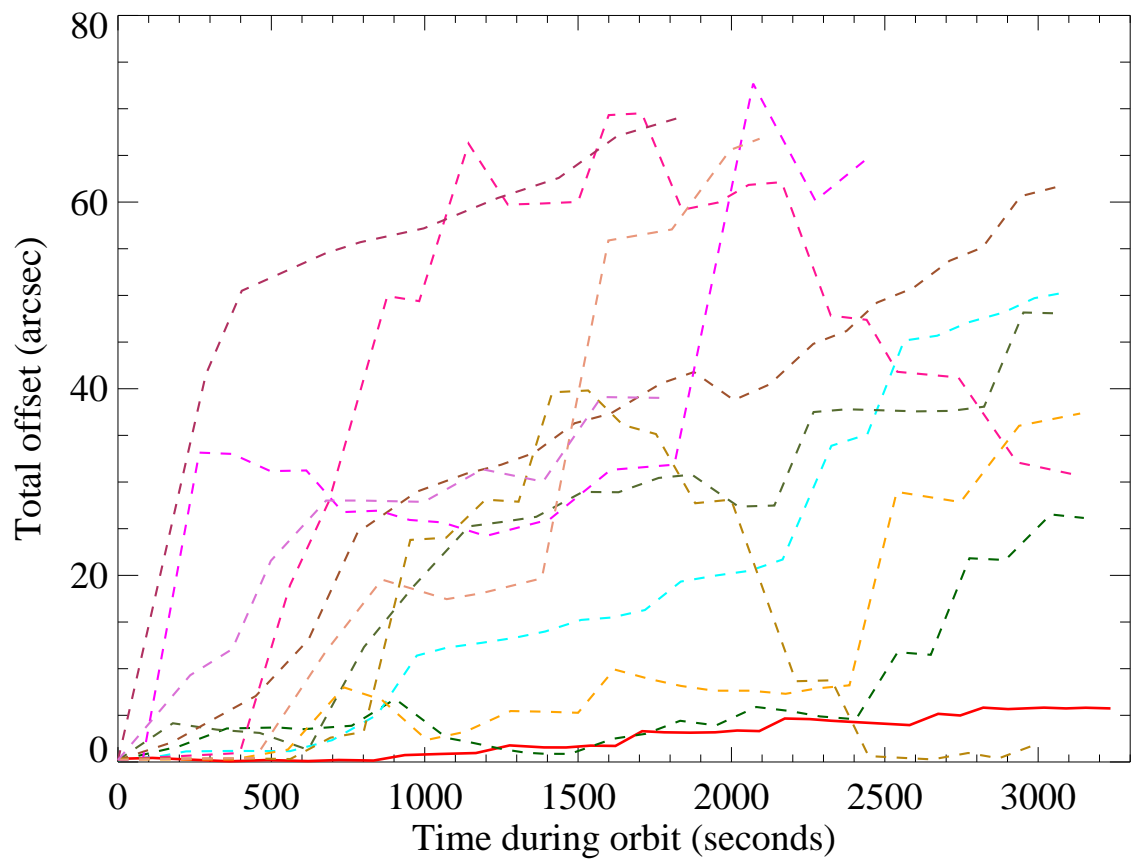}
\caption{\label{fg:outlier}  Gyro drift.  Each line shows the drift between the commanded position and the encountered in a single orbit under the ``DASH'' protocol of gyroscopic control.  The filled line shows the first result, near the expectation, and before the onset of degradation of gyro 2.  The others occurred during the usage of the degraded gyro 2.}
\end{figure}

\begin{figure}[ht]
\vspace*{150mm}
\figurenum{2}
\includegraphics{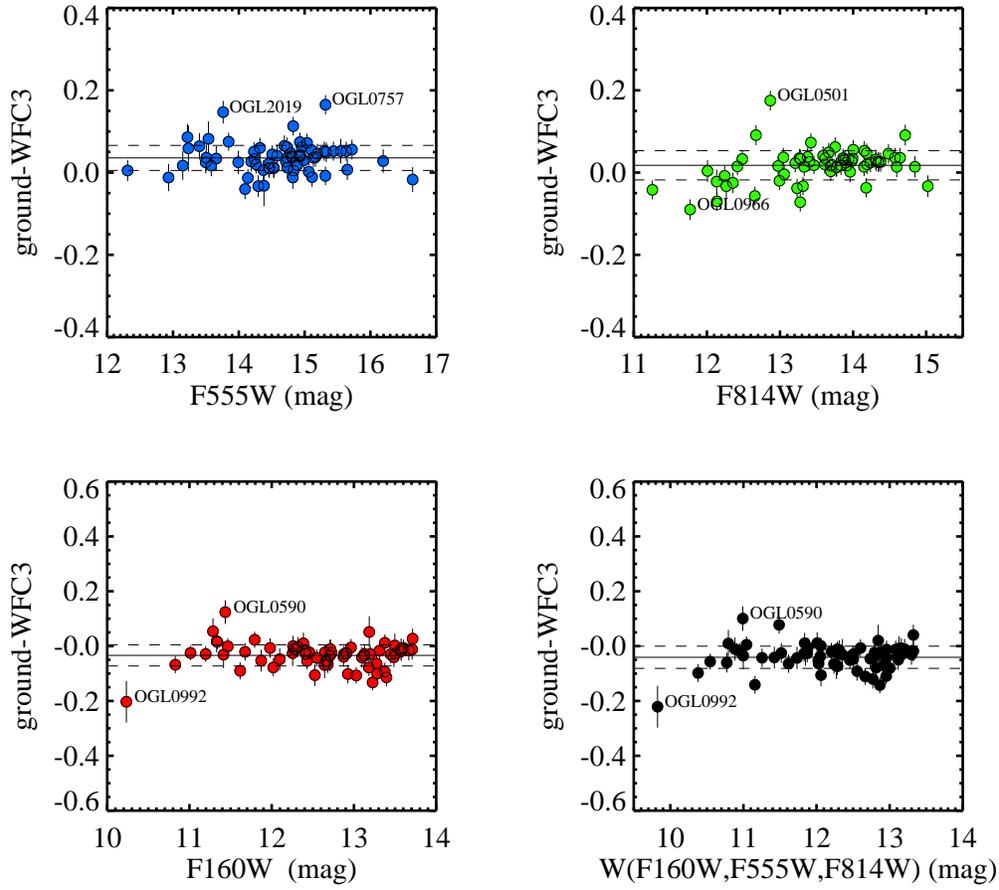}
\caption{\label{fg:outlier}  Comparison of Cepheid mean magnitudes in three {\it HST} WFC3 bands for observations obtained with {\it HST} and from the ground (transformed to the {\it HST} system). }
\end{figure}

\begin{figure}[ht]
\vspace*{200mm}
\figurenum{3}
\includegraphics{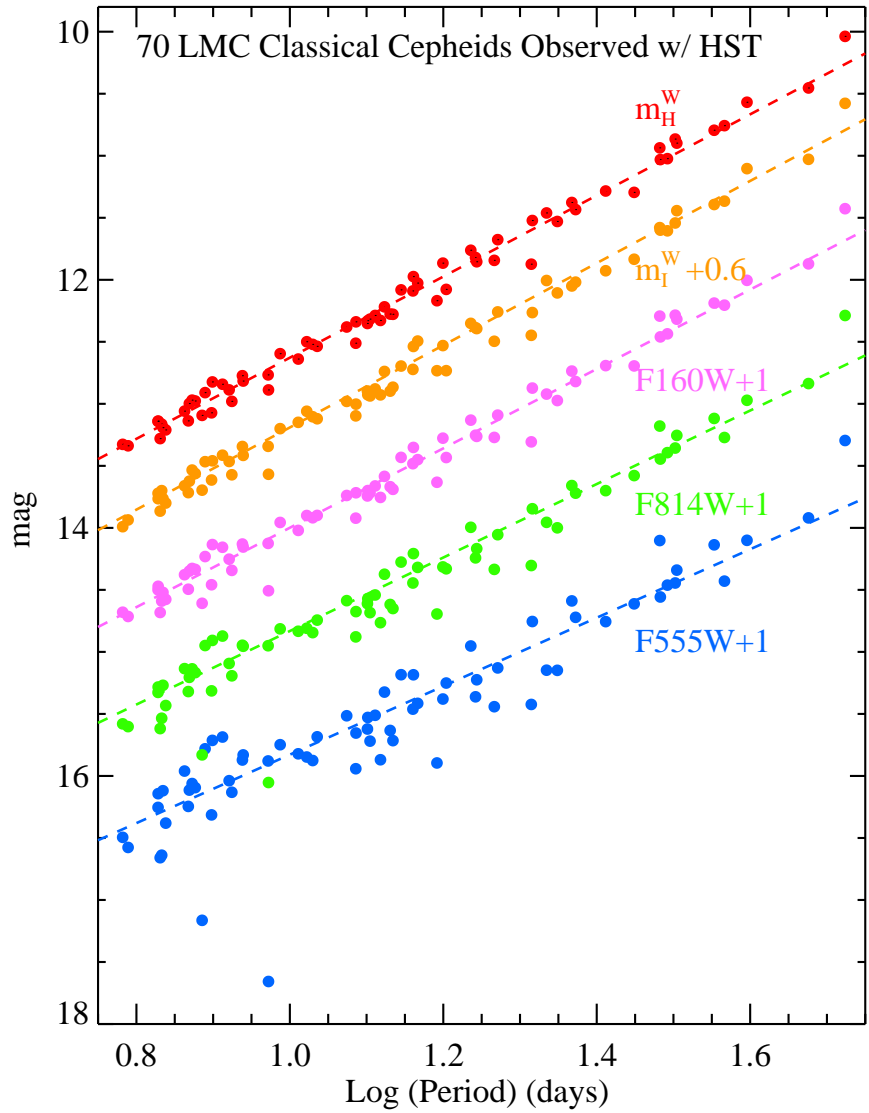}
\caption{\label{fg:outlier} Period-mean magnitude relation for the 70 LMC Cepheids with slopes and statistics given in Table 3.}
\end{figure}

\begin{figure}[ht]
\vspace*{200mm}
\figurenum{4}
\includegraphics{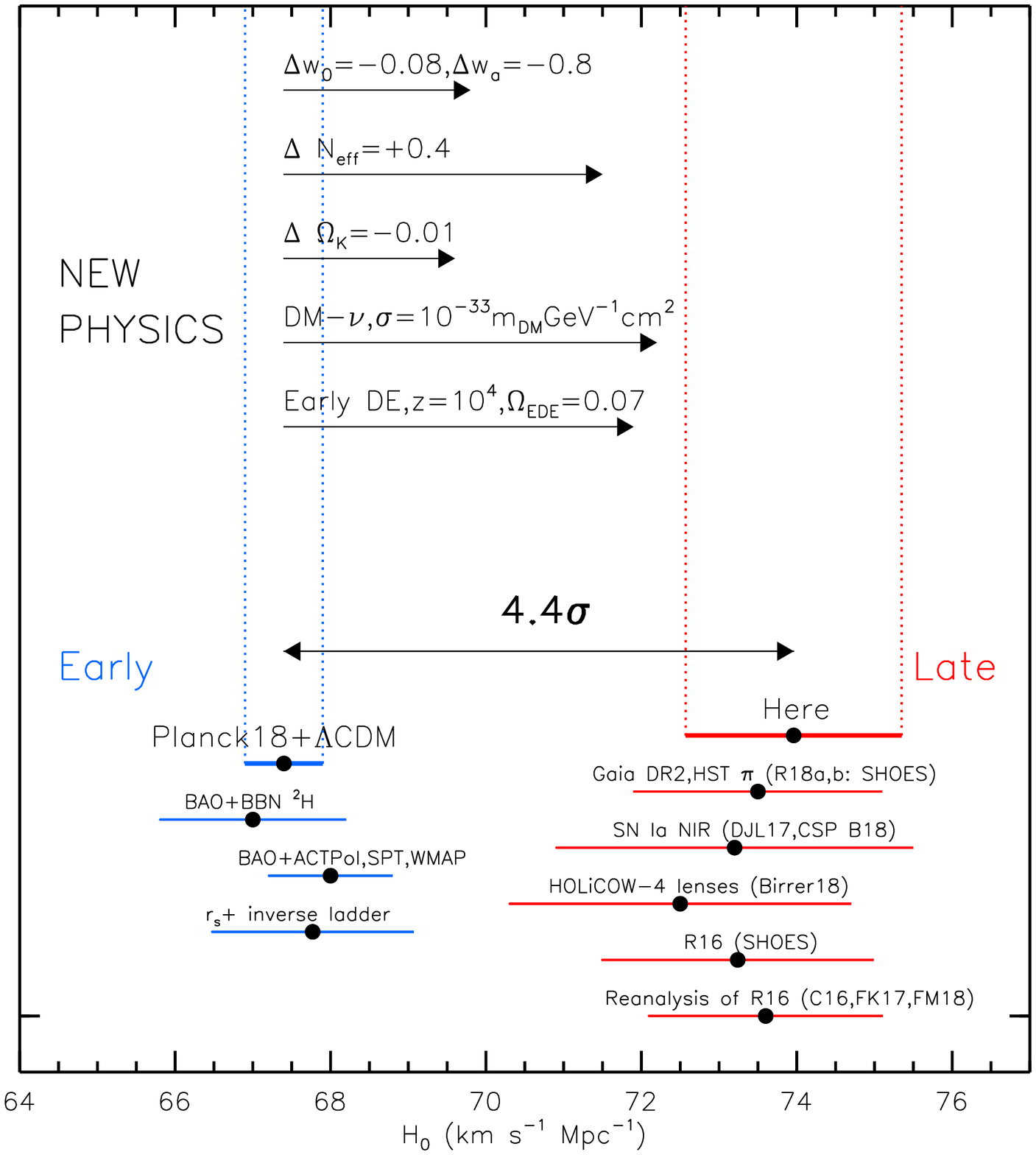}
\caption{\label{fg:outlier} The $4.4\sigma$
    difference between local measurements of H$_0$ and the value predicted
    from {\it Planck}+$\Lambda$CDM.  We show local results presented by Riess
    et al.~(2016), reanalysis by C16 \citep{Cardona:2017}, FK17 \citep{Follin:2017}, or FM18 \citep{Feeney:2017}, the HOLiCOW lensing results
    from Birrer18 \citep{Birrer:2018}, a replacement of optical SN data with NIR in DJL17 \citep{Dhawan:2017} and B18 \citep{Burns:2018}, and a revised geometric anchor from HST and {\it Gaia} DR2 parallaxes (R18a,b).  Other early universe scales are shown in blue.  Possible physics causes for a 2--4\% change in H$_0$ include time-dependent dark energy or nonzero curvature, while a larger 5--8\% difference may come from dark matter interaction, early dark energy or additional relativistic particles. }
\end{figure}

\begin{figure}[ht]
\vspace*{200mm}
\figurenum{5}
\includegraphics{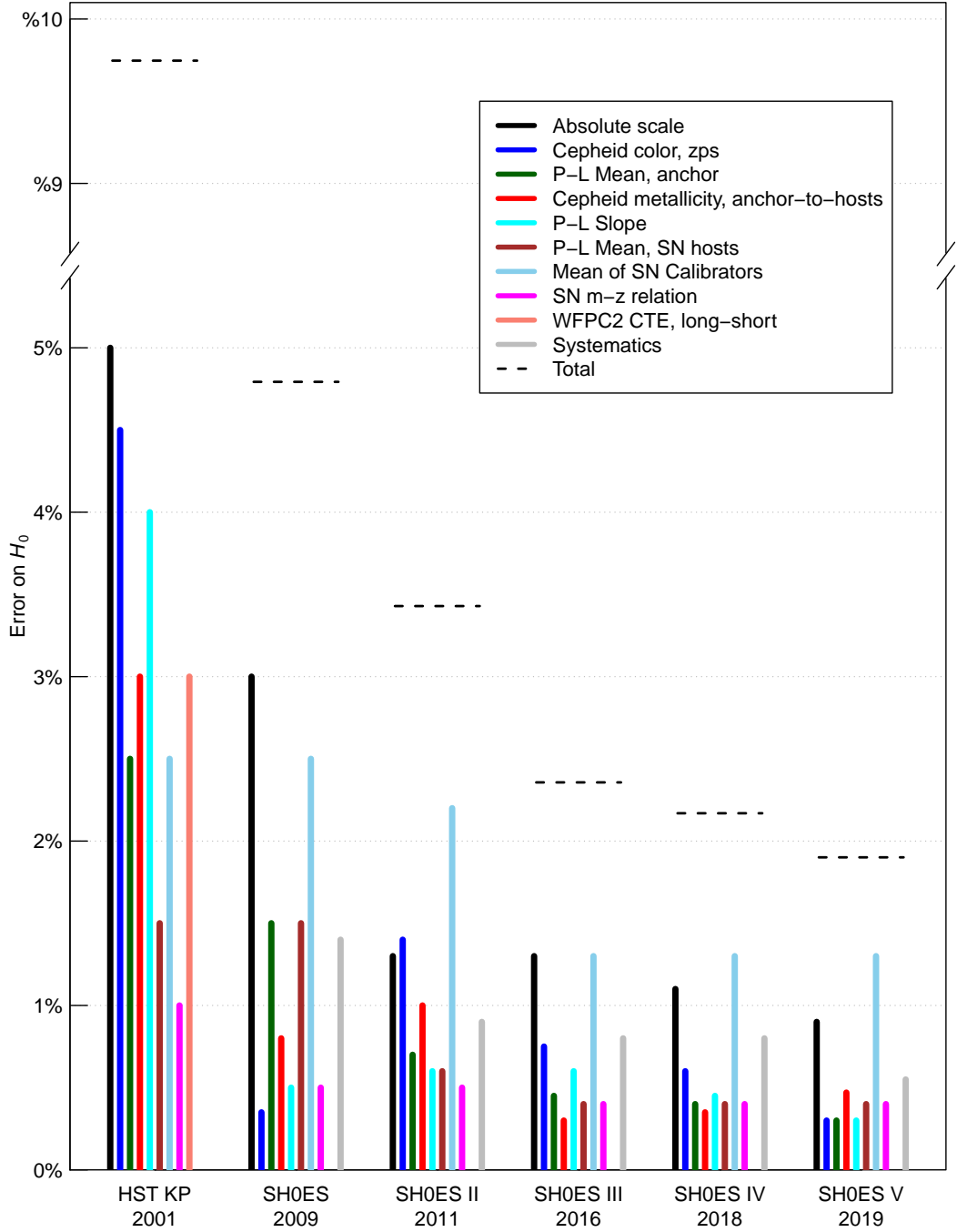}
\caption{\label{fg:outlier} Evolution of Error Budgets for SH0ES distance ladder and a comparison to the HST Key Project \citet{freedman01}.  Each error source (length of bar) contributes in quadrature (its square) to the total error, thus larger errors will be more dominant than the visual (linear) comparison suggests.}
\end{figure}

\vfill
\eject

\clearpage
\bibliographystyle{aasjournal} %
\bibliography{bibdesk}
\clearpage
\end{document}